# Partial-wave series expansions in spherical coordinates for the acoustic field of vortex beams generated from a finite circular aperture[†]

F.G. Mitri*

*Abstract* – Stemming from the Rayleigh-Sommerfeld surface integral, the addition theorems for the spherical wave and Legendre functions, and a weighing function describing the behavior of the radial component $v_{\rho_1}$ of the normal velocity at the surface of a finite circular radiating source, partial-wave series expansions are derived for the incident field of acoustic spiraling (vortex) beams in a spherical coordinate system centered on the axis of wave propagation. Examples for vortex beams, comprising $\rho$-vortex, zeroth-order and higher-order Bessel-Gauss and Bessel, truncated Neumann-Gauss and Hankel-Gauss, Laguerre-Gauss, and other Gaussian-type vortex beams are considered. The mathematical expressions are exact solutions of the Helmholtz equation. The results presented here are particularly useful to accurately evaluate analytically and compute numerically the acoustic scattering and other mechanical effects of finite vortex beams, such as the axial and 3D acoustic radiation force and torque components on a sphere of any (isotropic, anisotropic etc.) material (fluid, elastic, viscoelastic etc.). Numerical predictions allow optimal design of parameters in applications including but not limited to acoustical tweezers, acousto-fluidics, beam-forming design and imaging to name a few.

## I. INTRODUCTION

**P**ARTIAL-WAVE SERIES EXPANSIONS (PWSEs) are an important mathematical tool in various fields, such as optics, quantum mechanics and acoustics, which allow efficient modeling of the incident beam of propagating waves, and computation of the scattering, and other mechanical effects, such as the radiation force and torque on a particle or a cluster of particles.

In a spherical coordinates system, a weighted sum of spherical multipole waves, called a partial-wave series expansion (PWSE), is often used to model the incident beam. This method has been successfully applied in the study of the arbitrary acoustic scattering [1, 2], radiation force [3-5] and torque [6-8] of unbounded (infinite) vortex beams. Nevertheless, considering the fact that unbounded beams (or waves) carry an infinite amount of energy, and thus, are practically unrealizable, it is of some importance to further develop efficient tools to model the acoustics of finite vortex beams.

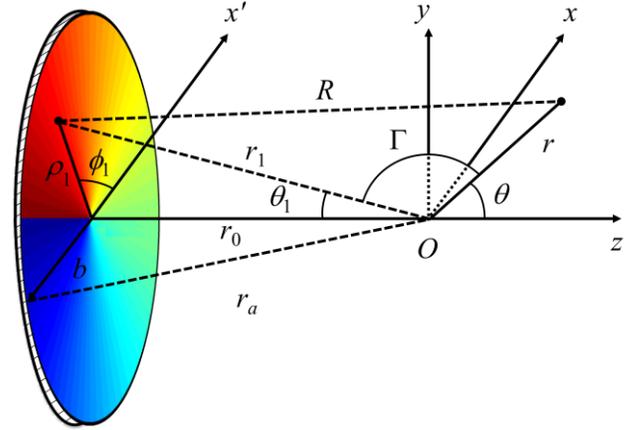

Fig. 1. Geometry of the problem used in the derivation of the incident acoustic fields of finite vortex beams. The color distribution on the surface of the circular aperture represents the phase variation, typical for vortex beams.

Previous works using the PWSE method have considered the finite (bounded) non-vortex beam case [9-16], whereas other studies have focused on the physical structure of rotating sound fields [17], such as a propeller [18], which can be useful in computing the noise from a rotor. At the same time, current research is investigating the scattering [1, 2, 19-22], radiation forces [3-5, 23, 24] and torques [6-8] on a spherical particle, in which infinite vortex (Bessel) beams were considered. However, acoustical vortex beams which carry an orbital angular momentum and emanate from a finite aperture are practically used in particle manipulation and rotation applications [25-28].

In this paper, a general method is presented for any finite (bounded) beam of vortex type (Fig. 1), as opposed to waves of infinite extent. The analysis is based on the Rayleigh surface integral [29] and the addition theorem for the spherical wave functions [30, 31] and the Legendre functions to derive appropriate PWSEs for the incident field describing the bounded vortex beam. Applications of the method for calculating the acoustic scattering, radiation force and torque from finite vortex beams on a sphere centered on or off the axis of wave propagation are discussed, and the appropriate equations are established.

## II. METHOD

Consider a finite circular piezo-disk transducer with a surface of radius $b$ in a non-viscous fluid medium. A spherical coordinates system $(r, \theta, \phi)$ is chosen to coincide with the center $O$ of the Cartesian coordinates system $(x, y, z)$ (Fig. 1). This choice simplifies the derivation of the scattering and other mechanical effects of vortex beams for a sphere centered on the beam's axis at the point $O$, given in Section IV, and the off-axial case, discussed in Section V. The coordinates systems are situated at an axial distance $r_0$ from the center of the circular transducer. The description of the incident acoustic field produced by the finite source, and represented by its scalar velocity potential field $\Phi_i$, is obtained from the Rayleigh-Sommerfeld integral as [29],

$$\Phi_i = \frac{1}{2\pi} \iint_{S_r} \frac{v(\rho_1, \phi_1)\big|_{z=0} e^{i(kR-\omega t)}}{R} dS_r, \quad (1)$$

where $R$ is the distance from the observation point to the finite source of circular surface $S_r$, the parameter $k$ is the wave number of the acoustic radiation, $\omega$ is the angular frequency, and $v$ is the normal velocity at $z = 0$. In the case of a uniform vibrational surface, $v$ is unitary [10-12, 32, 33]. However, it can be expressed by different mathematical functions whether the radiator is simply supported [32-34], clamped [32, 33], described by a rotating (propeller) [18] or a monopole ring source [35], apodized by a Gaussian distribution [9, 33], or excited according to its radially-symmetric vibrational modes [15] describing zero-order Bessel beams [36, 37].

For the purpose of the present study, apodizations of the normal velocity at $z = 0$ describing vortex beams are considered, for which the normal velocity is a separable function expressed as,

$$v(\rho_1, \phi_1)\big|_{z=0} = v_{\rho_1} e^{im\phi_1}, \quad (2)$$

where $\rho_1$ is the radial distance from the center of the radiator to a point on its flat surface, the parameter $m \neq 0$ is a positive or negative integer describing the topological charge (or order) of the beam, and $\phi_1$ is the azimuthal angle at the surface of the finite source (Fig. 1). Eq.(2) indicates that the normal velocity is separable into a radial and an azimuthal component, described by the functions $v_{\rho_1}$ and $e^{im\phi_1}$, respectively.

Suppressing the time dependence $e^{-i\omega t}$ for convenience, and using the addition theorem for the spherical functions, i.e. (10.1.45) and (10.1.46) in [31] such that $r \leq r_1$, (1) is expressed as,

$$\Phi_i = \frac{ik}{2\pi} \sum_{n=0}^{\infty} (2n+1) j_n(kr) \\ \times \iint_{S_r} v_{\rho_1} e^{im\phi_1} h_n^{(1)}(kr_1) P_n(\cos\Gamma) dS_r, \quad (3)$$

where $j_n(\cdot)$ and $h_n^{(1)}(\cdot)$ are the spherical Bessel and Hankel functions of the first kind, $P_n(\cdot)$ are the Legendre functions, and the differential surface $dS_r = \rho_1 d\rho_1 d\phi_1 = r_1 dr_1 d\phi_1$, since $r_1^2 = \rho_1^2 + r_0^2$.

Making use of the addition theorem for the Legendre functions [38, 39] using the definition of the angles as given in Fig. 1, $P_n(\cos\Gamma)$ can be expressed as ((3.19), p. 65 in [40]),

$$P_n(\cos\Gamma) = \sum_{\ell=0}^{n} (2-\delta_{\ell,0}) \frac{(n-\ell)!}{(n+\ell)!} \\ \times P_n^{\ell}(\cos(\pi-\theta)) P_n^{\ell}(\cos\theta_1) \cos\ell(\phi-\phi_1), \quad (4)$$

where $\delta_{ij}$ is the Kronecker delta function, and $P_n^{\ell}(\cdot)$ are the associated Legendre functions. Integrating both sides of (4) with respect to $\phi_1$ using the property of the following integral,

$$\int_0^{2\pi} e^{im\phi_1} \cos\ell(\phi-\phi_1) d\phi_1 = \begin{cases} 0, & \ell \neq m \\ 2\pi, & \ell = m = 0 \\ \pi e^{im\phi}, & \ell = m \neq 0 \end{cases} \quad (5)$$

gives for $\ell = m \neq 0$ after substituting (4) and (5) into (3), the incident velocity potential as,

$$\Phi_i = \frac{ie^{im\phi}}{k} \\ \times \sum_{n=|m|}^{\infty} \Lambda_{n,m} (-1)^{n+m} (2n+1) j_n(kr) \frac{(n-m)!}{(n+m)!} P_n^m(\cos\theta), \quad (6)$$

where the coefficient,

$$\Lambda_{n,m} = \int_{kr_0}^{kr_a} (kr_1) v_{\rho_1} h_n^{(1)}(kr_1) P_n^m\left(\frac{r_0}{r_1}\right) d(kr_1). \quad (7)$$

Eqs.(6),(7) describe the incident velocity potential for any vortex-type beam with its axis of wave propagation coinciding with the $z$-axis (Fig. 1). Depending on the form of apodizations of $v_{\rho_1}$, different types of beams can be obtained as follows. [Note that the single-integral in (7) can be evaluated numerically using the standard Gauss-Legendre quadrature or the trapezoidal method, as no closed-form expressions for the considered beams have been attained yet].

### A. $\rho$-vortex beam

Consider an acoustic vortex for which $v_{\rho_1}$ is expressed as,

$$v_{\rho_1} = v_{\rho_1}^{\rho-vortex} \\ = V_0 \rho_1^m \\ = V_0 \left(\sqrt{r_1^2 - r_0^2}\right)^m, \quad (8)$$

where $V_0$ is the velocity amplitude. Substituting (7) into (6) using (8), the incident velocity potential for an acoustic $\rho$-vortex beam with its axis of wave propagation coinciding with the $z$-axis can be obtained. If $m = 0$, the field corresponds to a circular disk with uniform vibration.

## B. Zeroth-order Bessel and Bessel-Gauss vortex beams (ZOBVB)

The standard lowest-order solution of Bessel beams, known as a zeroth-order Bessel beam [41, 42], is known to possess some properties such as the production of a beam profile with a long depth of field [43]. This particular solution is also of non-vortex type. Nevertheless, the possibility of creating a zeroth-order Bessel-like beam that spirals around the axis of wave propagation, called here ZOBVB, has been demonstrated in optics [44]. It is however important to note that the ZOBVB has an axial null, which is an intrinsic characteristic of a helicoidal beam. In the acoustic context, if $v_{\rho_1}$ is considered such that,

$$\begin{aligned} v_{\rho_1} &= v_{\rho_1}^{ZOBVB} \\ &= V_0 J_0(k_\rho \rho_1) \\ &= V_0 J_0\left(k_\rho \sqrt{r_1^2 - r_0^2}\right), \end{aligned} \quad (9)$$

where $k_\rho = k \sin\beta$, with $\beta$ corresponding to the half-cone angle of the beam, and $J_0(.)$ is the zero-order cylindrical Bessel function of the first kind, the beam can be of vortex nature.

Substituting (7) into (6) using (9), the incident velocity potential for a finite ZOBVB can be obtained.

Another variance of the ZOBVB is considered such that,

$$\begin{aligned} v_{\rho_1} &= v_{\rho_1}^{ZOBGVB} \\ &= V_0 J_0(k_\rho \rho_1) e^{-\rho_1^2/w_0^2}, \end{aligned} \quad (10)$$

where $w_0$ is the beam's waist (known also as the spot size). This corresponds to the radial component of the normal velocity for a zeroth-order Bessel-Gauss vortex beam (ZOBGVB).

For collimated waves such that $w_0 \to \infty$, (10) reduces to (9). Moreover, for a finite beam-waist $w_0 \neq 0$ and $m = 0$, the Bessel-Gauss beam is of *non-vortex* type, and its optical counterpart has been initially introduced in [45, 46].

## C. High-order Bessel-Gauss, Bessel, truncated Bessel, Neumann and Hankel vortex beams

Consider now an acoustic vortex for which the radial component of the normal velocity is expressed as,

$$\begin{aligned} v_{\rho_1} &= v_{\rho_1}^{HOBGVB} \\ &= V_0 J_m(k_\rho \rho_1) e^{-\rho_1^2/w_0^2}. \end{aligned} \quad (11)$$

Substituting (7) into (6) using (11), the incident velocity potential for a finite high-order Bessel-Gauss vortex beam (HOBGVB) is obtained.

When $w_0 \to \infty$, the incident velocity potential for a finite high-order Bessel vortex beam (HOBVB) is obtained. In this limit, and for the case where the half-cone angle $\beta (\to 0)$ is sufficiently small and the partial-wave number $n \to \infty$, the expression (9.1.71) in [31] can be used along with the equation relating the associated Legendre functions of negative order with those of positive order, (12.81) in [39], so that the cylindrical Bessel function may be approximated as,

$$J_m\left(k_\rho \sqrt{r_1^2 - r_0^2}\right) \approx n^m (-1)^m \frac{(n-m)!}{(n+m)!} P_n^m(\cos\beta)\bigg|_{\substack{n\to\infty \\ \beta\to 0}}, (12)$$

as long as [47], $\sqrt{r_1^2 - r_0^2} \approx (n+1/2)/k$. Note also that in the limit of infinite waves, such that $kr_a$ (or $kb$) $\to \infty$, an approximate value of the integral, obtained numerically, may be used such that,

$$\int_{kr_0}^{kr_a \to \infty} (kr_1) h_n^{(1)}(kr_1) P_n^m\left(\frac{r_0}{r_1}\right) d(kr_1) \approx i^{-n-m} n^{-m} \frac{(n+m)!}{(n-m)!} e^{ikr_0}. (13)$$

Therefore, substituting (7) into (6) using (11) as well as the approximations (12) and (13) leads to the incident velocity potential for a HOBVB of infinite extent, which is expressed as,

$$\Phi_i^{\infty, HOBVB} = \frac{iV_0}{k} e^{i(kr_0 + m\phi)} \\ \times \sum_{n=|m|}^{\infty} i^{n-m}(2n+1) j_n(kr) \frac{(n-m)!}{(n+m)!} P_n^m(\cos\theta) P_n^m(\cos\beta). \quad (14)$$

Note that an equivalent form of (14) was given in earlier works [19, 20] for the case of infinite HOBVBs.

Note here that both the finite and infinite zero-order Bessel (Gauss) non-vortex beam results follow immediately for $m = 0$ [15, 16]. Moreover, the case of a finite beam with uniform vibration at the source, as well as the case of infinite plane progressive waves can be recovered by setting $\beta = m = 0$.

It is important to emphasize also that for such vortex beams with $|m| > 0$, the incident field vanishes *on the axis* of wave propagation.

Other types of vortex beams with modified functions, such as the high-order truncated Bessel-Gauss, Neumann-Gauss and Hankel-Gauss beams may be obtained by replacing the regular cylindrical Bessel function $J_m(k_\rho \rho_1)$ in (11) by an appropriate truncated (denoted by an overbar) Bessel function of the first kind $\overline{J_m(k_\rho \rho_1)}$ (for which the zeroth-order had been investigated in optics [48]), Neumann (or Bessel function of the second kind) function $\overline{Y_m(k_\rho \rho_1)}$, or Hankel functions of the first $\overline{H_m^{(1)}(k_\rho \rho_1)}$ or second $\overline{H_m^{(2)}(k_\rho \rho_1)}$ kind, respectively. The truncated functions ensure continuity at $\rho_1 = 0$. Such truncation schemes have been previously applied for the optical zero-order Neumann [49, 50] and acoustical Hankel beams [51-53].

## D. Laguerre-Gaussian (LG) and elegant Laguerre-Gaussian (eLG) vortex beams

Earlier works [54, 55] in the field of acoustics suggested the description of *symmetric* (non-vortex) radiated field from ultrasonic transducers using Laguerre-Gaussian (LG) functions. Generally, LG beams are paraxial solutions to the wave equation, however, when generated from a finite circular piston transducer using the Rayleigh integral [55], the resulting beam is an exact solution of the Helmholtz equation. Here, an acoustic *asymmetric* LG vortex profile is considered such that the radial component of the normal velocity at the surface of a circular transducer ($z = 0$) is expressed as [56],

$$v_{\rho_1} = v_{\rho_1}^{LGVB}$$
$$= V_0 \left(\frac{\rho_1 \sqrt{2}}{w_0}\right)^m L_\eta^m \left(\frac{2\rho_1^2}{w_0^2}\right) e^{-\rho_1^2/w_0^2}, \quad (15)$$

where $L_\eta^m(.)$ are the associated Laguerre functions of degree $\eta$ and order $m$. The other parameters are given in the previous sections.

Substituting (7) into (6) with the use of (15), the incident velocity potential for a finite LG vortex beam is obtained.

It is interesting to note that the LG functions may be approximated by a cylindrical Bessel function of the first kind when the degree $\eta \to \infty$ (See section 8.22 in [57], [58, 59]). Moreover, research in optics suggested that the azimuthally-symmetric LG beams are special cases of high-order Bessel-Gauss beams and the equivalence between these two types of beams is established [60].

Consider now the eLG solution, which differs from the standard LG beam in that the latter is scaled by $\sqrt{2}$ at $z = 0$. Furthermore, in the paraxial approximation, eLG beams contain associated Laguerre functions with complex argument (during propagation, i.e. $z > 0$), whereas for the standard LG beams, the argument is real [56, 61]. The radial component of the normal velocity at the surface of a circular transducer ($z = 0$) for an eLG beam is expressed as [56],

$$v_{\rho_1} = v_{\rho_1}^{eLGVB}$$
$$= V_0 \left(\frac{\rho_1}{w_0}\right)^m L_\eta^m \left(\frac{\rho_1^2}{w_0^2}\right) e^{-\rho_1^2/w_0^2}, \quad (16)$$

so that the incident velocity potential is determined by substituting (7) into (6) with the use of (16).

The connection of the eLG beams with the Bessel-Gaussian and the Bessel solutions has been established [62] in the optical context, and it can be directly applied in the acoustics context in a similar approach.

### E. Other Gaussian-type vortex beams

Gaussian beams have been largely used in various ultrasonic applications including acoustic levitation, ultrasonic microscopy, high-intensity focused ultrasound (HIFU), and acoustical tweezers for particle manipulation, to name a few. Both focused [63] and unfocused beams [64] have been studied, which are paraxial solutions to the scalar wave equation; hence, they do not satisfy the Helmholtz equation. For a tightly focused wave-front, (i.e., the beam waist radius is of the same order as the wavelength), the beam profile no longer remains Gaussian, and often approximation schemes and higher-order corrections are needed to describe the propagating field with some degree of accuracy. Fortunately, the complex-source-point method provides an improved and more accurate modeling for strongly focused quasi-Gaussian beams [65], which exactly satisfy the Helmholtz equation. In the following, other Gaussian-type vortex beams are treated.

Consider an acoustic vortex for which the radial component of the normal velocity is expressed as,

$$v_{\rho_1} = v_{\rho_1}^{AVGB}$$
$$= V_0 \left(\frac{\rho_1}{w_0}\right)^{2\eta+m} e^{-\rho_1^2/w_0^2}, \quad (17)$$

where $\eta$ is defined as the degree of the beam. This type of beams has been introduced in optics and termed anomalous vortex (AV), since it exhibits an eLG profile in the far-field (or the focal plane) in free space [66]. Such a beam is characterized by a single-ringed intensity distribution in the transverse plane. Note that when $\eta = 0$ and $m \neq 0$, (17) will correspond to the radial component of the normal velocity profile of a Gaussian vortex beam. When $m = 0$ and $\eta \neq 0$, (17) reduces to the radial component of the normal velocity profile of a hollow (non-vortex) Gaussian beam [67], which has been extensively investigated in optics in possible atom guiding, trapping and focusing applications. Finally, when $\eta = m = 0$, (17) reduces to the radial component of the normal velocity profile of the fundamental Gaussian beam.

Substituting (7) into (6) using (17), the incident velocity potential for a finite acoustical AV Gaussian beam is obtained.

The analysis can be also extended to other types of Gaussian beams, such as the acoustical sinh-Gaussian vortex beam (sGVB) with degree $\eta$ and order $m$, for which the zero-order optical counterpart has been investigated in the paraxial limit [68] (one may also replace the hyperbolic function "sinh" by the standard sine function to produce other types of beams). The normal velocity for a sGVB is expressed as,

$$v_{\rho_1} = v_{\rho_1}^{sGVB}$$
$$= V_0 \left[\sinh\left(\frac{\rho_1}{w_0}\right)\right]^\eta e^{-\rho_1^2/w_0^2}, \quad (18)$$

so that the incident velocity potential for a finite acoustical sGVB is obtained by inserting (7) into (6) using (18).

There exist additional optical Gaussian-type vortex beams, such as the Hermite-Gauss, the Hermite-sinusoidal-Gaussian, the Hypergeometric, the Hypergeometric-Gaussian, the asymmetric Bessel-Gaussian that may be derived from the asymmetric Bessel mode [69], the Bessel-Gaussian of fractional type $\alpha$ [22, 70] (and others), for which some of their acoustical counterparts were briefly reviewed in [71]. The present analysis can be directly extended to include such beams in a similar procedure.

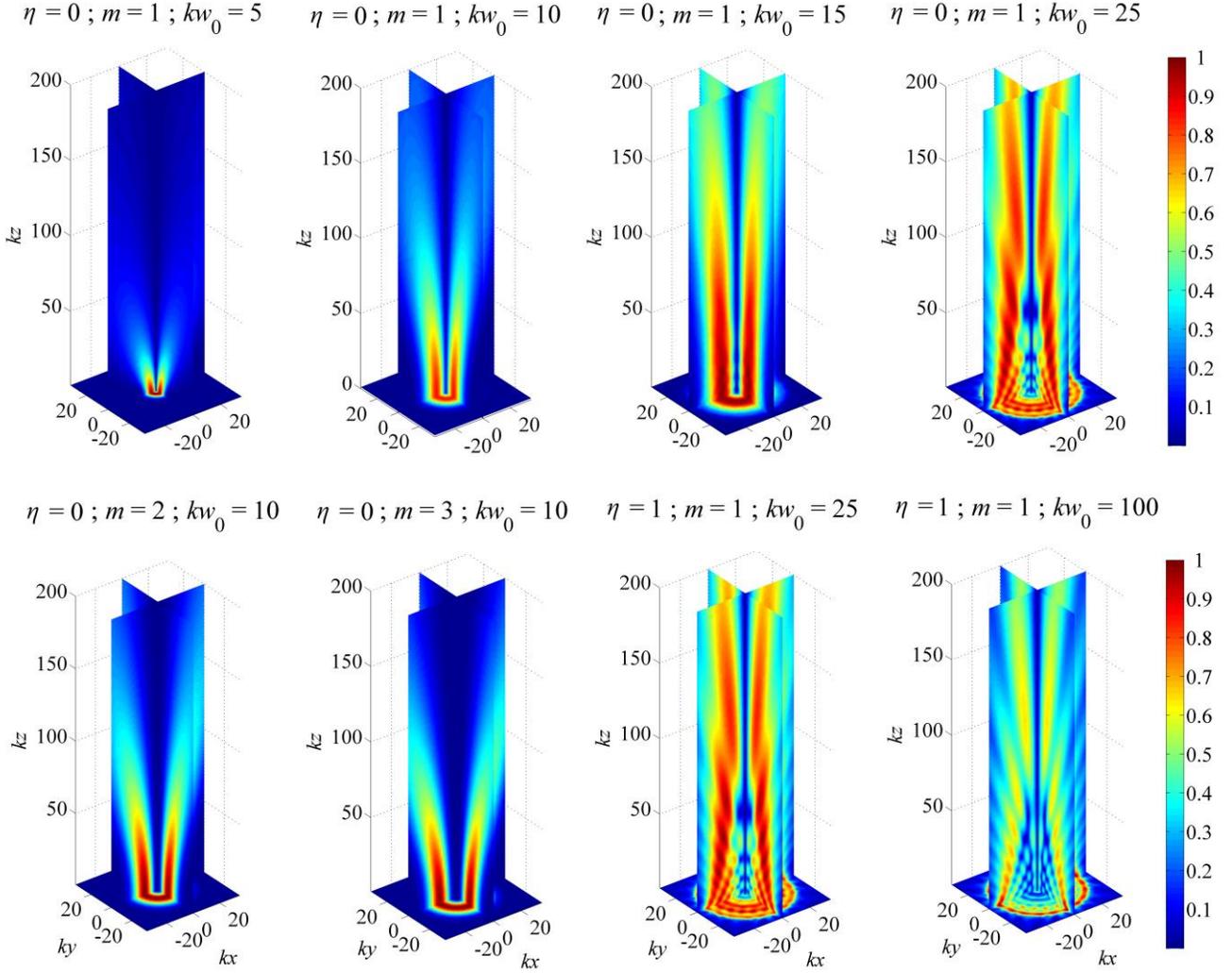

Fig. 2. Normalized magnitudes (to unity) of the incident velocity potential representing the propagated field of a Laguerre-Gaussian vortex beam for $kb = 30$, $m \geq 1$, $\eta \geq 0$, and for different values of the dimensionless waist $kw_0$. The numerical computations were performed in the transverse dimensionless ranges $-35 \leq (kx, ky) \leq 35$, and the axial dimensionless range $1 < kz \leq 200$. The transverse magnitude plots were evaluated at $kz = 1$.

## III. BEAMFORMING SIMULATIONS – EXAMPLE FOR (HIGH-ORDER) LAGUERRE-GAUSSIAN VORTEX BEAMS

As an example to illustrate the analysis, numerical simulations for the acoustic field of a finite LGVB for $kb = 30$ and $m \geq 1$ in nonviscous water are performed by computational evaluation of the magnitude of the steady-state incident velocity potential field, given by (6) and using (7) and (15). Emphasis is given on the degree $\eta$, the order $m$, the dimensionless beam waist $kw_0$, and the dimensionless radius of the circular radiator $kb$. The series in (6) are truncated such that the criterion for convergence, $\left|\Phi_{n,i} - \Phi_{n-1,i}\right| / \left|\Phi_{n-1,i}\right|$ is as small as $<< \varepsilon \sim 10^{-6}$, where $\Phi_{n,i}$ is the incident velocity potential estimate obtained by truncation of the PWSE after the $n$th term. Eq.(7) is evaluated by numerical integration using the trapezoidal method with a sampling step of $10^{-5}$, which ensures adequate convergence.

The first row in Fig. 2 shows the effect of varying the dimensionless beam waist for a unit vortex LGVB with zeroth degree ($\eta = 0$). The beam becomes collimated as $kw_0$ increases, and the *axial* velocity potential (or pressure) magnitude vanishes at the center of the beam. The effect of changing the order $m$ is also analyzed by comparing the second panel ($\eta = 0$; $m = 1$; $kw_0 = 10$) in the first row with those displayed in the first ($\eta = 0$; $m = 2$; $kw_0 = 10$) and second ($\eta = 0$; $m = 3$; $kw_0 = 10$) panels in the second row of Fig. 2. As the order increases from $m = 1$ to $m = 3$, the radius/diameter of the central null increases. In addition, increasing the degree $\eta$ of the beam does not seem to induce a significant effect on the velocity potential magnitude plots by comparing panel four in the first row with panel three in the second row of Fig. 2. Finally, the effect of increasing $kw_0$ for a first-order LGVB with unitary degree is shown in the fourth panel, second row of Fig. 2. The collimation of the beam appears to be more enhanced when $kw_0$ increases and $\to \infty$.

## IV. APPLICATION FOR EVALUATING THE AXIAL SCATTERING, RADIATION FORCE AND TORQUE ON A SPHERE – EXAMPLE FOR A (HIGH-ORDER) LAGUERRE-GAUSSIAN VORTEX BEAM

When a sphere is centered on the beam's axis, defined as the axial case, the partial-wave series expansions (PWSEs) provided earlier can be used to advantage in evaluating the scattering, radiation force and torque induced by vortex beams. The generalization of the (resonance) scattering theory by an elastic sphere (or spherical shell) taking into account the geometry of the beam in acoustics [2, 72] has demonstrated the importance of the PWSE method in modeling any beam of arbitrary shape using the so-called beam-shape coefficients (BSCs). The BSCs describe the beam's characteristics in the spherical coordinate system, can be used to reconstruct the incident beam field, and are defined independently of the presence and size of the scatterer.

The BSCs, denoted by $a_{p,q}$, can be expressed as [1],

$$a_{p,q} = \frac{1}{j_p(kr)\Phi_0} \int_{\phi=0}^{2\pi}\left[\int_{\theta=0}^{\pi}\Phi_i(r,\theta,\phi)Y_p^{q*}(\theta,\phi)\sin\theta d\theta\right]d\phi, \quad (19)$$

where $\Phi_0$ is defined here as, $\Phi_0 = iV_0/k$,

$$Y_p^q(\theta,\phi) = \sqrt{\frac{(2p+1)}{4\pi}\frac{(p-q)!}{(p+q)!}}P_p^q(\cos\theta)e^{iq\phi}, \text{ are the}$$

Laplace spherical harmonics, and the superscript * denotes a complex conjugate.

Substituting the incident velocity potential field (given by (6) for each beam type using the PWSE method) into (19) and performing the angular integration by applying the orthogonality condition of the spherical harmonics ((12.154) in [39]), the expression for the BSCs reduce to a simplified form (on-axis). Those coefficients are thus termed *axial* BSCs.

As an example, the axial BSCs for a finite LGVB are derived after substituting (7) into (6) using (15), and performing the appropriate algebraic manipulation, such that, the final result is given by,

$$a_{n,m}^{axial\,LGVB} = \Lambda_{n,m,\eta}^{LGVB}(-1)^{n+m}\sqrt{4\pi(2n+1)\frac{(n-m)!}{(n+m)!}}\delta_{pn}\delta_{qm}, \quad (20)$$

where,

$$\Lambda_{n,m,\eta}^{LGVB} = \int_{kr_0}^{kr_a}\left\{\begin{array}{l}(kr_1)\left(\frac{\sqrt{2(r_1^2-r_0^2)}}{w_0}\right)^m L_\eta^m\left(\frac{2(r_1^2-r_0^2)}{w_0^2}\right)e^{-(r_1^2-r_0^2)/w_0^2}\\ \times h_n^{(1)}(kr_1)P_n^m\left(\frac{r_0}{r_1}\right)\end{array}\right\}d(kr_1). \quad (21)$$

Now that the axial BSCs are determined, evaluation of the axial scattering, acoustic radiation force and torque becomes possible. Applying the generalized formalisms for the scattering [1, 2, 73, 74], radiation force [3, 4, 75] and torque [8] of arbitrary beams on a sphere, the axial components of the scattering form function $f_\infty^{axial}$, the radiation force function $Y_z^{axial}$, and the dimensionless torque $\tau_z^{axial}$ are expressed, respectively, as

$$f_\infty^{axial\,LGVB} = \frac{2}{ika}\sum_{n=|m|}^{\infty}i^{-n}a_{n,m}^{axial\,LGVB}S_nY_n^m(\theta,\phi), \quad (22)$$

$$Y_z^{axial\,LGVB} = \frac{1}{\pi(ka)^2}\text{Im}\sum_{n=|m|}^{\infty}\left\{\begin{array}{l}\sqrt{\frac{(n-m+1)(n+m+1)}{(2n+1)(2n+3)}}\\ \times\Gamma_n a_{n,m}^{axial\,LGVB}a_{n+1,m}^{axial\,LGVB*}\end{array}\right\}, \quad (23)$$

$$\tau_z^{axial\,LGVB} = -\frac{m}{\pi(ka)^3}\text{Re}\left\{\sum_{n=|m|}^{\infty}\left|a_{n,m}^{axial\,LGVB}\right|^2 S_n^*(1+S_n)\right\}, \quad (24)$$

where Re and Im denote the real and imaginary parts of a complex number, respectively, $S_n$ are the scattering coefficients determined by applying specific boundary condition depending on the physical properties of the sphere, and the parameter $\Gamma_n = S_n + S_{n+1}^* + 2S_n S_{n+1}^*$. Note that the sphere has to be viscoelastic so as to experience a non-zero torque [6, 7]. Moreover, the axial radiation force function (23) can be further simplified for vortex beams centered on a sphere by introducing the real $\alpha_n$ and imaginary $\beta_n$ parts of the scattering coefficients $S_n$, such that,

$$Y_z^{axial} = \frac{1}{\pi(ka)^2}$$
$$\times\sum_{n=|m|}^{\infty}\sqrt{\frac{(n-m+1)(n+m+1)}{(2n+1)(2n+3)}}\left\{\begin{array}{l}\text{Re}\left[a_{n,m}^{axial}a_{n+1,m}^{axial*}\right]\left[\beta_n(1+2\alpha_{n+1})-\beta_{n+1}(1+2\alpha_n)\right]\\ +\text{Im}\left[a_{n,m}^{axial}a_{n+1,m}^{axial*}\right]\left[\alpha_n+\alpha_{n+1}+2(\alpha_n\alpha_{n+1}+\beta_n\beta_{n+1})\right]\end{array}\right\}, \quad (25)$$

which is a generalization of (26) in [76] for the case of (axial) asymmetric vortex beams.

## V. THE OFF-AXIAL CASE

For the case where the sphere is shifted off-axially, the present analysis based on PWSE solutions can be used in determining the off-axial scattering, radiation force and torque, with appropriate transformations of the spherical wave functions using the translational (or rotational) addition theorem. Recent works have investigated the off-axial resonance scattering from a sphere in the field of a finite circular transducer [72], and the radiation force on a sphere in the field of a spherically-focused transducer [77] using this method. These transformations applied to the axial BSCs produce a PWSE for the off-axial BSCs, which can be used to compute the off-axial scattering, radiation force and torque, respectively.

The expression for the off-axial BSCs in terms of the axial BSCs using the translational addition theorem of the spherical wave functions is given by [72, 77, 78],

$$a_{p,q} = \sum_{n=0}^{\infty} \sum_{m=-n}^{n} a_{n,m}^{axial} \hat{S}_{np}^{mq}(k\mathbf{d}), \qquad (26)$$

where $\hat{S}_{np}^{mq}(k\mathbf{d})$ are the translation coefficients (given explicitly by (3.81) in [40]), and $\mathbf{d}$ is the translation displacement vector (i.e., off-axial displacement vector shift).

Based on the procedure described in Section IV, the axial BSCs, expressed using a PWSE for any finite beam, can be substituted into (26) to obtain the off-axial BSCs in the translated coordinate system, and then compute the off-axial scattering [1, 2, 73, 74], and the 3D components of the radiation force [3, 4, 75] and torque [8] of arbitrary beams on a sphere.

## VI. CONCLUSIONS

A rigorous method based on the PWSE for modeling the acoustic field of finite vortex (asymmetric) beams is developed, for the prediction of time-harmonic fields around circular disks. The method uses the Rayleigh surface integral and the addition theorem for the spherical wave functions and Legendre functions. Appropriate PWSEs for the incident field describing the bounded vortex beam are derived for a variety of beams. Moreover, applications of the method for calculating the acoustic scattering, radiation force and torque from finite vortex beams on a sphere centered on or off the axis of wave propagation are discussed, and the appropriate equations are established.

It should also be mentioned that the previous analyses in which high-order trigonometric (*non-vortex*) beams of infinite extent were considered [74, 79, 80], will benefit from the results of this study by replacing the phase factor $\exp(im\phi)$ by $\cos(m\phi)$ in the expressions for the incident fields.